\documentclass[prd,nofootinbib,preprint,superscriptaddress]{revtex4}
\pdfoutput=1
\usepackage[T1]{fontenc}
\usepackage{amsmath,amssymb}
\usepackage{epsfig}
\usepackage{graphicx}
\usepackage[usenames,dvipsnames]{color}
\usepackage{subfigure}
\usepackage{slashed}
\usepackage[colorlinks,citecolor=blue]{hyperref}
\usepackage{pdfpages}
\usepackage{color}
\usepackage{comment}

\begin{document}
\title{LIGO-VIRGO constraints on dark matter and leptogenesis triggered by a first order phase transition at high scale}

\author{Debasish Borah}
\email{dborah@iitg.ac.in}
\affiliation{Department of Physics, Indian Institute of Technology Guwahati, Assam 781039, India}

\author{Arnab Dasgupta}
\email{arnabdasgupta@pitt.edu}
\affiliation{Pittsburgh Particle Physics, Astrophysics, and Cosmology Center, Department of Physics and Astronomy, University of Pittsburgh, Pittsburgh, PA 15206, USA}

\author{Indrajit Saha}
\email{s.indrajit@iitg.ac.in}
\affiliation{Department of Physics, Indian Institute of Technology Guwahati, Assam 781039, India}

\begin{abstract}
We study the possibility of constraining a scenario with high scale first order phase transition (FOPT) responsible for the cogenesis of baryon and dark matter using gravitational wave (GW) (non)-observations. While the FOPT at high scale is responsible for generating baryon asymmetry through leptogenesis and dark matter via the \textit{mass-gain} mechanism, the resulting GW spectrum falls within the ongoing LIGO-VIRGO experimental sensitivity. The dark matter is preferred to be in the non-thermal ballpark with sub-GeV masses and the criteria of successful dark matter relic rules out a large portion of the parameter space consistent with high scale FOPT and successful leptogenesis. Some part of the parameter space allowed from dark matter and leptogenesis criteria also gives rise to a large signal-to-noise ratio at ongoing experiments and hence can be disfavoured in a conservative way from the non-observation of such stochastic GW background. Future data from ongoing and planned experiments will offer a complementary and indirect probe of the remaining parameter space which is typically outside the reach of any direct experimental probe. 
\end{abstract}

\maketitle

\section{Introduction}
Presence of dark matter (DM) and baryon asymmetry in the universe (BAU) has been suggested by several astrophysical and cosmological observations \cite{Zyla:2020zbs, Aghanim:2018eyx}. These have been two longstanding problems in particle physics and cosmology given the fact that the standard model (SM) of particle physics fails to provide any explanations for the same. While several beyond standard model (BSM) proposals have been put forward to explain these observed phenomena, none of them have been experimentally tested yet. With popular production mechanisms for BAU namely, baryogenesis \cite{Weinberg:1979bt, Kolb:1979qa} or leptogenesis \cite{Fukugita:1986hr} typically remain a high scale phenomena out of direct reach of terrestrial experiments, particle DM has not been discovered yet at direct detection experiments \cite{LUX-ZEPLIN:2022qhg}. This has motivated alternative and indirect ways of probing such mechanisms behind the origin of DM and BAU. One such avenue is the detection of stochastic gravitational wave (GW) background, which has been utilised in several baryogenesis or leptogenesis scenarios \cite{Hall:2019ank, Dror:2019syi, Blasi:2020wpy, Fornal:2020esl, Samanta:2020cdk, Barman:2022yos, Baldes:2021vyz, Azatov:2021irb, Huang:2022vkf, Dasgupta:2022isg, Barman:2022pdo, Datta:2022tab, Borah:2022cdx} as well as particle DM models \cite{Hall:2019rld, Yuan:2021ebu, Tsukada:2020lgt, Chatrchyan:2020pzh, Bian:2021vmi, Samanta:2021mdm, Borah:2022byb, Azatov:2021ifm, Azatov:2022tii, Baldes:2022oev, Borah:2022iym, Borah:2022vsu, Shibuya:2022xkj}.

While most of the previous works focused on future detection of stochastic GW arising in DM and baryogenesis setups, here we consider the possibility of constraining such models with existing data from LIGO-VIRGO-KAGRA (LVK) experiments taken during their first three observing runs (O1, O2, O3). There have already been tight constraints on isotropic GW background from LVK observations \cite{KAGRA:2021kbb}. The same constraints have been used in the context of particle physics models with first order phase transition (FOPT) capable of generating stochastic GW in the LVK ballpark \cite{Romero:2021kby, Jiang:2022mzt, Badger:2022nwo, Huang:2021rrk}. Motivated by this, here we consider a high scale leptogenesis and DM triggered by a FOPT. Similar to the baryogenesis and leptogenesis scenarios proposed in \cite{Baldes:2021vyz, Huang:2022vkf, Dasgupta:2022isg}, we consider a minimal setup where both DM and right handed neutrino (RHN) responsible for leptogenesis acquire masses in a FOPT by crossing the relativistic bubble walls. In our previous work \cite{Borah:2022cdx}, we focused on a low scale FOPT or low scale leptogenesis such that the resulting GW spectrum remains within the sensitivities of future experiments. In the present work, we consider a high scale version of this setup which can already be constrained by existing GW experiments like LIGO-VIRGO. We show the parameter space consistent with successful leptogenesis which is disfavoured by LVK data. The minimal version of this model also predicts non-thermal fermion singlet DM with mass in the sub-GeV ballpark.

This paper is organised as follows. In section \ref{sec1} we briefly discuss our model followed by the details of our results in section \ref{sec2}. We finally conclude in section \ref{sec:conclude}.

\section{The Model}
\label{sec1}
In order to show the key results, we consider a minimal setup where the type-I seesaw for light neutrino masses \cite{Minkowski:1977sc,Yanagida:1979as,Gell-Mann:1979vob,Mohapatra:1980yp} is extended by a scalar doublet $\eta$ with an additional $Z_2$ symmetry under which $\eta$ and one of the RHNs are odd. We also impose a classical conformal invariance such that the mass terms arise only after a singlet scalar $S$ acquires a non-zero vacuum expectation value (VEV) while also driving a FOPT. The relevant part of the Yukawa Lagrangian is given by 
\begin{equation}\label{IRHYukawa}
{\cal L} \ \supset \ \frac{1}{2}Y'_{ij} S N_i N_j + \left(Y_{\alpha 1} \, \bar{L}_\alpha \tilde{\eta} N_1  + \sum_{j=2,3} (y_D)_{\alpha j} \, \bar{L}_\alpha \tilde{\Phi_1} N_j  + \text{h.c.} \right) \ . 
\end{equation}
where $\Phi_1$ is the SM Higgs doublet. Thus, two of the RHNs even under $Z_2$ take part in type-I seesaw while the $Z_2$-odd sector contributes radiatively to one of the light neutrino masses in scotogenic fashion \cite{Ma:2006km}. The scalar potential of the model can be written as
\begin{align}
V(\Phi_1,\eta, S) & \ = \ \frac{\lambda_1}{4}|\Phi_1|^4+\frac{\lambda_2}{4}|\eta|^4+\lambda_3|\Phi_1|^2|\eta|^2 + \frac{1}{4} \lambda_S S^4 +\lambda_4|\Phi_1^\dag \eta|^2 + \left[\frac{\lambda_5}{2}(\Phi_1^\dag \eta)^2 + \text{h.c.}\right] \nonumber \\
& \qquad +\lambda_6|\Phi_1|^2 S^2 + \lambda_7 |\eta|^2 S^2 \, . \label{eq:tree potential}
\end{align}
Since we are interested in the singlet scalar induced FOPT at high scale, we denote the singlet scalar as $S=(\phi+M)/\sqrt{2}$ with $M$ denoting the singlet scalar VEV as well as the scale of renormalisation. Since we are assuming a classical conformal invariance, the singlet scalar VEV not only decides the physical masses of RHNs and $\eta$, but also generates the scale of electroweak symmetry breaking dynamically. This constraints the parameter $\lambda_6 <0$ to a small value for high scale FOPT. For simplicity, we consider the two $Z_2$-even RHNs to be quasi-degenerate while the $Z_2$-odd RHN to be much lighter, playing the role of DM.

\section{Results and Discussion}
\label{sec2}

In order to study the details of the FOPT, we consider the tree level potential $V_{\rm tree}$ mentioned above, one-loop Coleman-Weinberg potential $V_{\rm CW}$\cite{Coleman:1973jx} along with the finite-temperature potential $V_{\rm th}$ \cite{Dolan:1973qd,Quiros:1999jp} such that the full potential is $V_{\rm tot} = V_{\rm tree} + V_{\rm CW} +V_{\rm th}$. While calculating the thermal potential we also include the Daisy corrections \cite{Fendley:1987ef,Parwani:1991gq,Arnold:1992rz} which improve the perturbative expansion during the FOPT. Out of the two popularly used schemes namely, Parwani method and Arnold-Espinosa method, we use the latter. The details of the finite-temperature potential can be found in appendix \ref{appen1}. The FOPT proceeds via tunneling, the rate for which is estimated by calculating the bounce solution. The bounce action $S_3 (T)$ determines the tunneling rate per unit volume defined as
\begin{align}
\Gamma (T) = \mathcal{A}(T) e^{-S_3(T)/T}.
\end{align}
where $\mathcal{A}(T)\sim T^4\left( \frac{S_3(T)}{2\pi T}\right)^{3/2}$ and $S_3(T)$ are respectively determined by the dimensional analysis and given by the classical configuration, called bounce. The bounce solution can then be obtained by following the prescription given in \cite{Linde:1980tt}. The details of this prescription and our calculation are given in appendix \ref{appen3}. The nucleation temperature $T_n$ of the FOPT is then calculated by comparing the tunnelling rate to the Hubble expansion rate as
\begin{align}
    \Gamma (T_n) = {\bf H}^4(T_n) = {\bf H^4_*},
\end{align}
with ${\bf H}_* \equiv {\bf H}(T=T_n)$. The strength of the FOPT is conventionally decided by the order parameter $\phi(T_c) / T_c \equiv v_c/T_c$ with $\phi(T_c)$ being the singlet scalar VEV at critical temperature $T=T_c$ at which the two minima of the potential are degenerate. Larger is the order parameter $v_c/T_c >1$, stronger is the FOPT. The bounce calculation is done by rewriting the zero temperature one-loop effective potential as \cite{Jinno:2016knw, Iso:2009nw}
\begin{align}
     V_{0} &= V_{\rm tree} + V_{\rm CW}, \nonumber \\
    &= \frac{1}{4} \lambda_S(t) G^4(t) \phi^4
\end{align}
where $t={\rm log}(\phi/\mu)$ with $\mu=M$ being the scale of renormalisation and the function $G(t)$ is given by
\begin{align}
    G(t) = e^{-\int^t_0 dt' \gamma(t')},\; \gamma(t) = \frac{1}{32\pi^2} {\rm Tr}[Y'^{\dagger} Y'].
\end{align}
The relevant Yukawa and scalar potential couplings as a function of energy scale are calculated by solving the corresponding renormalisation group evolution (RGE) equations \cite{Borah:2022cdx}, the details of which can be found in appendix \ref{appen2}. Finally, the temperature at which the FOPT is completed, known as the percolation temperature $T_p$ is calculated by following the prescription given in \cite{Ellis:2018mja, Ellis:2020nnr}. According to this prescription, $T_p$ is obtained from the probability of finding a point which is still in the false vacuum, given by
\begin{align}
    \mathcal{P}(T) = e^{-\mathcal{I}(T)}. \nonumber
\end{align}  
Here,
\begin{align}
    \mathcal{I}(T) = \frac{4\pi}{3}\int^{T_c}_T \frac{dT'}{T'^4}\frac{\Gamma(T')}{{\bf H}(T')}\left(\int^{T'}_T \frac{d\tilde{T}}{{\bf H}(\tilde{T})}\right)^3.
\end{align}
The percolation temperature is then calculated by using  $\mathcal{I}(T_p) = 0.34$ \cite{Ellis:2018mja} which implies that at least $34\%$ of the comoving volume is occupied by the true vacuum.


\begin{figure}[h]
    \centering
    \includegraphics[scale=0.5]{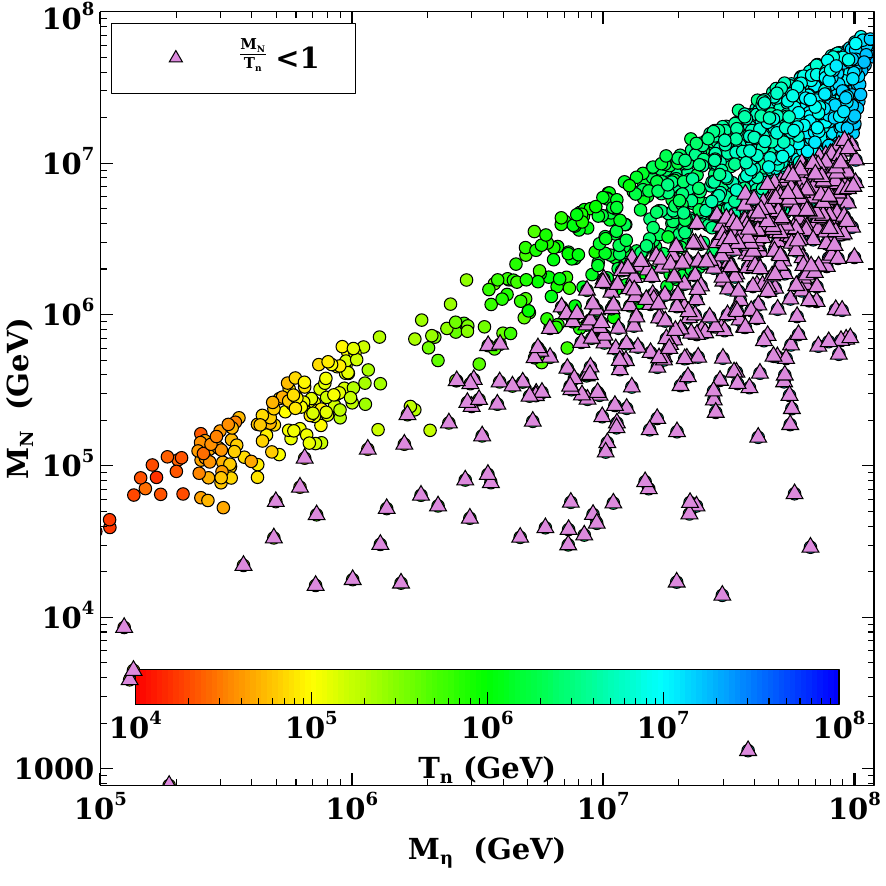}
       \includegraphics[scale=0.5]{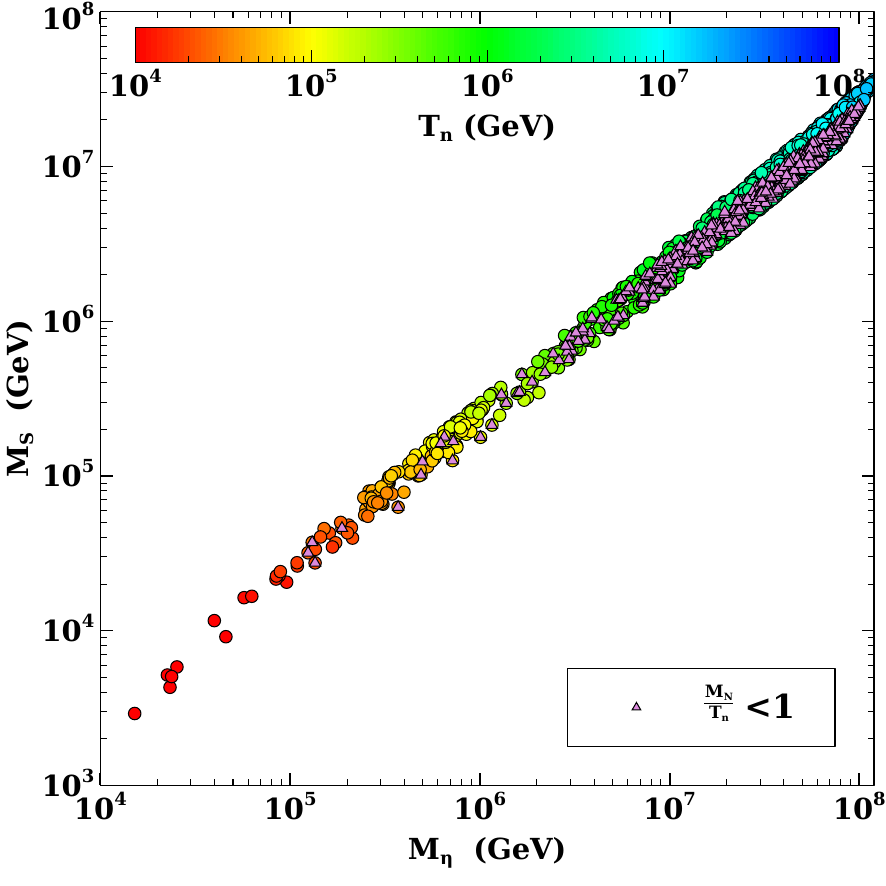}
    \caption{Parameter space in $M_\eta$ versus $M_N$ plane (left panel) and $M_\eta$ versus $M_S$ plane (right panel) consistent with a high scale FOPT $v_c/T_c >1$ with the colour code showing the corresponding nucleation temperature $T_n$. In this scan, the scale of phase transition (M) is varied from $10^4$ to $10^8$ GeV, $\lambda_7(0)$ is varied from 1 to 3, $Y'_{22}\sim Y'_{33}$ is varied from 0 to 1 and $Y'_{11}$ is varied upto $10^{-2}$.}
    \label{fig:1}
\end{figure}

In Fig. \ref{fig:1}, we show the parameter space in terms of physical masses of the heavier quasi-degenerate RHNs denoted by $M_N$, neutral real component of scalar doublet $\eta$ and scalar singlet denoted by $M_N, M_\eta, M_S$ respectively. The parameter space is consistent with a high scale FOPT with the colour code showing the nucleation temperature $T_n$. We also identify the points with $M_N < T_n$ where one has to consider thermal production of heavy neutrinos into account while estimating the lepton asymmetry produced.

In order to calculate the stochastic GW spectrum as a result of the FOPT, we take all relevant contributions into account from the bubble collisions~\cite{Turner:1990rc,Kosowsky:1991ua,Kosowsky:1992rz,Kosowsky:1992vn,Turner:1992tz}, the sound wave of the plasma~\cite{Hindmarsh:2013xza,Giblin:2014qia,Hindmarsh:2015qta,Hindmarsh:2017gnf} and the turbulence of the plasma~\cite{Kamionkowski:1993fg,Kosowsky:2001xp,Caprini:2006jb,Gogoberidze:2007an,Caprini:2009yp,Niksa:2018ofa}. The two important quantities namely, the duration of the phase transition and the latent heat released are calculated and parametrised in terms of $\frac{\beta}{{\bf H}(T)} \simeq T\frac{d}{dT} \left(\frac{S_3}{T} \right)$ and $\alpha_*$ respectively. The action is evaluated numerically by fitting our potential using the procedure laid out in \cite{Adams:1993zs} and utilised in our earlier work \cite{Borah:2022cdx}. The bubble wall velocity $v_w$ is estimated by first calculating the Jouguet velocity $v_J = \frac{1/\sqrt{3} + \sqrt{\alpha^2_* + 2\alpha_*/3}}{1+\alpha_*}$ \cite{Kamionkowski:1993fg, Steinhardt:1981ct, Espinosa:2010hh} which then leads to the bubble wall velocity $v_w$ as \cite{Lewicki:2021pgr}\footnote{See \cite{Ai:2023see} for a recent model-independent determination of wall velocity.}

\begin{equation}
    v_w = 
    \begin{cases}
    \sqrt{\frac{\Delta V_{\rm tot}}{\alpha_* \rho_{\rm rad}}} & \text{if} \,\, \sqrt{\frac{\Delta V_{\rm tot}}{\alpha_* \rho_{\rm rad}}} < v_J \\
    1 & \text{if}  \,\, \sqrt{\frac{\Delta V_{\rm tot}}{\alpha_* \rho_{\rm rad}}} \geq v_J. \\
    \end{cases}
\end{equation}
Here, $\rho_{\rm rad}=g_*\pi^2 T^4/30$ denotes the radiation energy density and $\Delta V_{\rm tot}$ denotes the energy difference between the true and the false vacua, given by
\begin{equation}
    \Delta V_{\rm tot} \equiv V_{\rm tot}(\phi_{\rm false},T)- V_{\rm tot}(\phi_{\rm true},T).
\end{equation}
The latent heat released during the phase transition can be estimated as
\begin{align}
    \alpha_* =\frac{\epsilon_*}{\rho_{\rm rad}},
    \label{alphastar}
\end{align}
with
\begin{align}
    \epsilon_* = \left[\Delta V_{\rm tot} - \frac{T}{4} \frac{\partial \Delta V_{\rm tot}}{\partial T}\right]_{T=T_n},
\end{align}
which is also related to the change in the trace of the energy-momentum tensor across the bubble wall \cite{Caprini:2019egz,Borah:2020wut}.

\begin{figure}[h]
    \centering
    \includegraphics[scale=0.6]{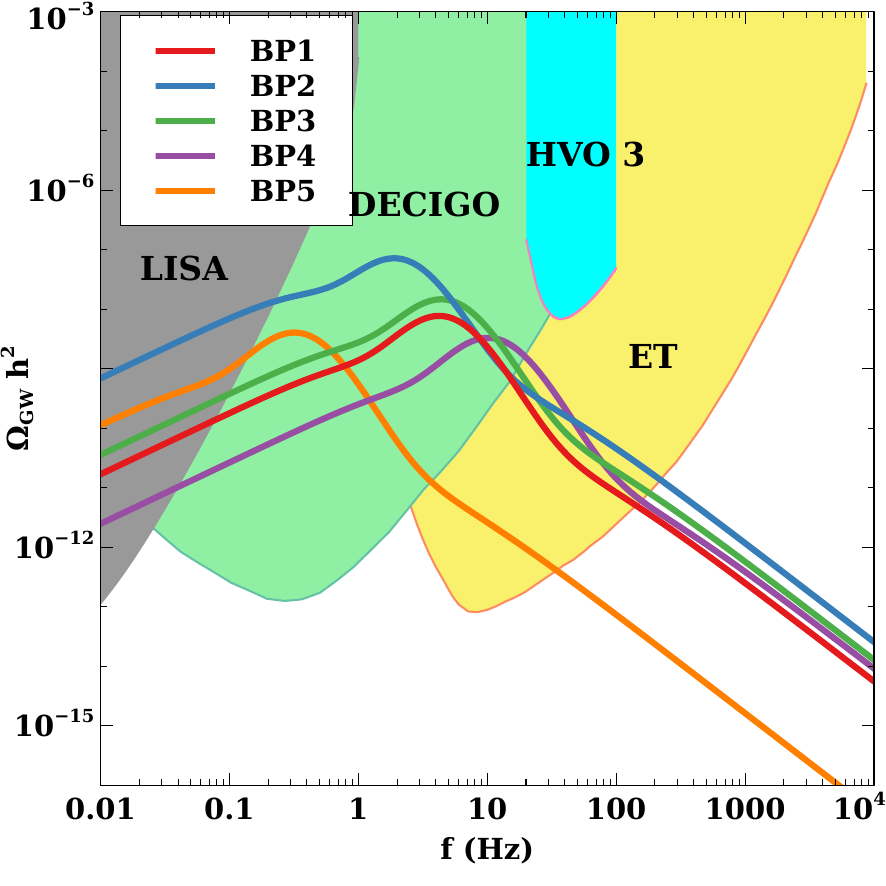}
    \caption{GW spectrum for different benchmark points of our model shown in table \ref{tab1}.}
    \label{fig:2}
\end{figure}

The stochastic GW spectrum, considering all the sources, can be written as \cite{Caprini:2015zlo,Caprini:2024hue}
\begin{align}
		\Omega_{\rm GW}(f) &= \Omega_\phi(f) + \Omega_{\rm sw}(f) + \Omega_{\rm turb}(f),
\end{align}
where $\Omega_\phi, \Omega_{\rm sw}, \Omega_{\rm turb}$ correspond to the individual contributions from bubble collisions, sound wave of the plasma and turbulence in the plasma respectively. In general, each of these contributions has a peak type feature with peak frequency $f_{\rm peak}$.  The spectrum can be parametrised as
\begin{align}
		\Omega h^2 (f) &= \mathcal{R}\Delta(v_w)\left(\frac{\kappa \alpha_*}{1+\alpha_*}\right)^p\left(\frac{{\bf H_*}}{\beta}\right)^q\mathcal{S}(f/f_{\rm peak})
	\end{align}
 where the pre-factor $\mathcal{R}$ takes into account the red-shift of the GW energy density, $\mathcal{S}(f/f_{\rm peak})$ parametrises the shape of the spectrum and $\Delta(v_w)$ is the normalization factor which depends on the bubble wall velocity $v_w$. $\kappa$ is the efficiency parameter which denotes the fraction of latent heat driving that particular source of GW in a first order phase transition. The numerical values of the parameters $p,q$ depend upon the source. For bubble collision as the source, the spectrum has been re-estimated in several recent works \cite{Ellis:2018mja, Ellis:2020nnr, Lewicki:2022pdb, Lewicki:2020jiv, Lewicki:2020azd} and can be written as \cite{Lewicki:2022pdb, Athron:2023xlk, Caprini:2024hue}
	\begin{equation}
\Omega_\phi h^2 = 1.67 \times 10^{-5} \left ( \frac{100}{g_*} \right)^{1/3} \left(\frac{{\bf H}_*}{\beta}\right)^2 \left(\frac{\kappa_\phi \alpha_*}{1+\alpha_*}\right)^2 \frac{A (a+b)^c}{\left[ b (f/f^\phi_{\rm peak})^{-a/c}+ a(f/f^\phi_{\rm peak})^{b/c} \right]^c} ,
	\end{equation}
where\footnote{The values of these parameters can change if the phase transition leads to breaking of gauge symmetry \cite{Lewicki:2022pdb}. In our setup, the singlet scalar induced phase transition does not lead to any gauge symmetry breaking which is consistent with the chosen values of $a,b,c$.}, $a=1.03, b=1.84, c=1.45$ and A=5.93$\times10^{-2}$ and the peak frequency being \cite{Lewicki:2022pdb, Athron:2023xlk, Caprini:2024hue}
	\begin{equation}
 f^\phi_{\rm peak} = 1.65 \times 10^{-5} {\rm Hz} \left ( \frac{g_*}{100} \right)^{1/6} \left ( \frac{T_n}{100 \; {\rm GeV}} \right ) \frac{0.64}{2\pi} \left(\frac{\beta}{{\bf H}_*}\right).
	\end{equation}
 The efficiency factor $\kappa_\phi$ for bubble collision is given by \cite{Kamionkowski:1993fg}
 \begin{equation}
     \kappa_\phi =\frac{1}{1+0.715\alpha_*}\left( 0.715 \alpha_* + \frac{4}{27} \sqrt{\frac{3\alpha_*}{2}} \right).
 \end{equation}
	The GW spectrum generated from the sound wave in the plasma has been studied through large hydrodynamical simulations \cite{Hindmarsh:2017gnf} which has also been updated in several recent works \cite{Caprini:2019egz, Guo:2020grp, Athron:2023xlk}. The corresponding spectrum can be written as \cite{Athron:2023xlk} 
	\begin{equation}
		\Omega_{\rm sw}h^2 = 2.59\times 10^{-6} \left(\frac{100}{g_*}\right)^{1/3}\left(\frac{\bf H_*}{\beta}\right) \left( \frac{\kappa_{\rm sw} \alpha_*}{1+\alpha_*}\right)^2 v_w\frac{7^{3.5}(f/f^{\rm sw}_{\rm peak})^3}{(4+3(f/f^{\rm sw}_{\rm peak})^2)^{3.5}} \Upsilon.
	\end{equation}
	The corresponding peak frequency is given by
	\begin{equation}
		f^{\rm sw}_{\rm peak}=8.9\times10^{-6}{\rm Hz} \left ( \frac{g_*}{100} \right)^{1/6} \frac{1}{v_w} \left ( \frac{T_n}{100 \; {\rm GeV}} \right )  \left(\frac{\beta}{{\bf H}_*}\right)(\frac{z_p}{10}).
	\end{equation}
 The efficiency factor for sound waves, applicable for relativistic bubble wall velocity $v_w \sim 1$ in our model, is \cite{Espinosa:2010hh}
 \begin{equation}
     \kappa_{\rm sw} =\frac{\alpha_*}{0.73+0.083 \sqrt{\alpha_*}+ \alpha_*}.
 \end{equation}
 Here, $\Upsilon=1-\frac{1}{\sqrt{1+2\tau_{sw}H_*}}$ is a suppression factor which depends on the lifetime of sound wave $\tau_{\rm sw}$\cite{Guo:2020grp} and it can be written as $\tau_{\rm sw}\sim R_*/\bar{U}_f$ with mean bubble separation, $R_*=(8\pi)^{1/3}v_w \beta$ and rms fluid velocity, $\bar{U}_f=\sqrt{3\kappa_{sw}\alpha_*/4(1+\alpha_*)}$; $z_p\sim 10$.
	Finally, the spectrum generated by the turbulence in the plasma is given by \cite{Caprini:2015zlo, Athron:2023xlk, Caprini:2024hue}
	\begin{equation}
		\Omega_{\rm turb}h^2 = 3.35\times 10^{-4} \left(\frac{100}{g_*}\right)^{1/3}\left(\frac{\bf H_*}{\beta}\right) \left( \frac{\kappa_{\rm turb} \alpha_*}{1+\alpha_*}\right)^{1.5}v_w \frac{(f/f^{\rm turb}_{\rm peak})^3}{(1+f/f^{\rm turb}_{\rm peak})^{3.6}(1+8\pi f/h_*)}
	\end{equation}
   with the peak frequency being \cite{Caprini:2015zlo}
	\begin{equation}
		f^{\rm turb}_{\rm peak}=2.7\times10^{-5}{\rm Hz} \left ( \frac{g_*}{100} \right)^{1/6} \frac{1}{v_w} \left ( \frac{T_n}{100 \; {\rm GeV}} \right )  \left(\frac{\beta}{{\bf H}_*}\right).
	\end{equation}
 The efficiency factor for turbulence is $\kappa_{\rm turb} \simeq 0.1 \kappa_{\rm sw}$ \cite{Caprini:2015zlo} and the  inverse Hubble time at the epoch of GW production, redshifted to today is
	\begin{equation}
		h_*=1.65\times10^{-5}{\rm Hz} \left ( \frac{g_*}{100} \right)^{1/6} \left ( \frac{T_n}{100 \; {\rm GeV}} \right ).
	\end{equation}
The total contribution to the stochastic GW spectrum is shown in Fig. \ref{fig:2} for a few benchmark points shown in table \ref{tab1}. Since we are focusing on high scale FOPT, we are showing the relevant sensitivities of future experiments like DECIGO \cite{Kawamura:2006up}, ET\,\cite{Punturo_2010} and ongoing LIGO-VIRGO (HVO3) \cite{LIGOScientific:2014pky, KAGRA:2021kbb} as shaded regions of different colours.

We then implement the leptogenesis via relativistic bubble wall or mass-gain mechanism \cite{Baldes:2021vyz, Huang:2022vkf, Dasgupta:2022isg, Borah:2022cdx} for the high scale FOPT scenario. The right handed neutrinos $N_{1,2,3}$ and scalar doublet $\eta$ acquire masses after entering the bubble formed due to the FOPT induced by the singlet scalar discussed above. The quasi-degenerate heavier RHNs namely, $N_{2,3}$ then decay into leptons and Higgs to generate the lepton asymmetry while $N_1$, being the lightest $Z_2$-odd particle emerges as the DM candidate. The CP asymmetry parameter corresponding to the CP violating decay of RHN $N_i$ (summing over all lepton flavours) is given by \cite{Pilaftsis:2003gt}
\begin{eqnarray}
\epsilon_{i} & = & \dfrac{\Gamma_{(N_{i}\longrightarrow \sum_{\alpha} L_{\alpha} \Phi_1 )}-\Gamma_{(N_{i}\longrightarrow \sum_{\alpha} L_{\alpha}^{c}\Phi^*_1)}}{\Gamma_{(N_{i}\longrightarrow \sum_{\alpha}L_{\alpha}\Phi_1)}+\Gamma_{(N_{i}\longrightarrow \sum_{i}L_{\alpha}^{c} \Phi^*_1)}}  \\ 
 & = &\dfrac{{\rm Im}[(y^{\dagger}_D y_D)_{ij}^{2}]}{(y^{\dagger}_D y_D)_{ii}(y^{\dagger}_D y_D)_{jj}}\dfrac{(M_{i}^{2}-M_{j}^{2})M_{i}\Gamma_{j}}{(M_{i}^{2}-M_{j}^{2})^{2}+M_{i}^{2}\Gamma_{j}^{2}}. \label{eq:asymmparameter}
\end{eqnarray}
where, only self-energy correction is considered, which dominate for quasi-degenerate RHNs. Here $M_i$ denotes the mass of RHN $N_i$. Since we consider quasi-degenerate $N_{2,3}$, we denote their mass as $M_N$ hereafter. The relativistic nature of the bubble walls arising out of the supercooled phase transition ensures the penetration of RHNs to maintain a large abundance inside the bubble, same as the equilibrium abundance without a Boltzmann suppression. For this, we need to ensure that the Lorentz boost of the bubble wall should be more than the Lorentz factor of the particle in the plasma frame 
\begin{align}
\gamma_w >\gamma_N \sim \frac{M_{N}}{T_n}
    \label{eq:gammaw}
\end{align}
where $T_n$ is the nucleation temperature. However, due to $T_n < M_{N}$, RHNs can not be thermally produced but yet having a large comoving abundance $Y_N=n_N/s$ inside the bubble without any Boltzmann suppression, with $n_N, s$ being equilibrium number density of $N$ and entropy density of the universe respectively. The comoving abundance of $N$ inside the bubble is then evaluated as
\begin{align}
    Y_N &= \frac{135}{8\pi^4}\xi(3)\frac{g_N}{g_*}
\end{align}
where $g_N$ and $g_*$ are the degrees of freedom of RHN $N$ and the total relativistic degrees of freedom of the universe, respectively. For the parameter space with $M_S > 2M_N$, we also take the additional contribution to $Y_N$ from singlet scalar $S$ decay. The final baryonic asymmetry can then be approximated as
\begin{align}
    Y_B =\frac{n_B-n_{\bar{B}}}{s}= \epsilon_N \kappa_{\rm sph}Y_N\left(\frac{T_n}{T_{\rm RH}}\right)^3.
    \label{eq:eps}
\end{align}
where $\epsilon_N \equiv \epsilon_{2,3} \simeq \sin(2\delta)/(16\pi)$ \cite{Pilaftsis:1998pd, Pilaftsis:2003gt} is the CP-asymmetry and $\delta$ is the relative CP phase between the quasi-degenerate RHNs, $\kappa_{\rm sph} = 8/23$ is the sphaleron conversion factor for our model, and $T_{\rm RH}$ is the reheating temperature after the FOPT. $T_{\rm RH}$ is defined as $T_{\rm RH} = {\rm Max}[T_n, T_{\rm inf}]$ \cite{Baldes:2021vyz} where $T_{\rm inf}$ can be found by comparing radiation energy density to the energy released from the FOPT or equivalently $\Delta V_{\rm tot}$ namely,
\begin{equation}
    \frac{g_*\pi^2}{30}T^4_{\rm inf} = \Delta V_{\rm tot}.
\end{equation}
We also check the feasibility of RHN decay into $L, \Phi_1$ at the reheating temperature by considering the thermal masses of daughter particles. Since thermal masses of $L, \Phi_1$ scale as $T$, the requirement of $M_{\Phi_1} (T_{\rm RH}) + M_L (T_{\rm RH}) < M_N (T_{\rm RH})$ leads to a lower bound $M_N/T_{\rm RH} >0.77$ which is satisfied by our chosen model parameters. The final baryon asymmetry calculated this way is then compared with the observed one $Y_B^{\rm obs}=(8.61\pm 0.05)\times 10^{-11}$ \cite{Planck:2018vyg} and required CP asymmetry parameter is obtained. The Dirac Yukawa coupling, estimated from type-I seesaw formula by considering active neutrino mass in the order of 0.01 eV, is given by
    \begin{align}
    y_D \equiv \sum_\alpha y_{D_{1\alpha}} \sim  2.3\times 10^{-8}\left(\frac{M_N}{1 ~{\rm GeV} }\right)^{1/2} \, .
\end{align}
Table \ref{tab2} shows the required CP asymmetry along with other relevant parameters involved in calculating baryon asymmetry for the benchmark points discussed before.

The validity of baryon asymmetry estimate given by Eq. \eqref{eq:eps} also depends upon the assumption that the washout processes are absent or negligible. In particular, the dominant wash-out coming from inverse decay should be suppressed to validate $Y_B$ estimated in Eq. \eqref{eq:eps}. While this is true due to $M_N > T_n$, presence of $L, \Phi_1$ with energy more than the mean thermal energy can induce such washout. As the RHNs $N_{2,3}$ enter the bubble, they receive a boost in the plasma frame with energy $\gamma_N M_{N}$, where $\gamma_N$  $\sim M_{N}/T_n$. Consequently, the decay products $L, \Phi_1$ and $L^c, \Phi^*_1$ also appear to be boosted, with the energy of the order $\frac{M^2_{N}}{T_n}$. These boosted particles can interact with SM particles in the plasma of energy $T_n$ and go through cascade scattering with particles in the plasma while redistributing their energies. At the same time, these energetic particles can also lead to washout asymmetry through on-shell processes such as $L \, \Phi_1$ $ \rightarrow$  $L^c \, \Phi^*_1$ via $N_{2,3}$. With the simplest assumption of energy distribution, the energy of boosted particles in $n$-th step cascade scattering is $\frac{M_{N_{2,3}}^2}{2^n T_n}$. Following the procedure of \cite{Baldes:2021vyz,Huang:2022vkf}, the rate of the washout in the above process can be estimated as $\Gamma_{\rm wash}\approx 2^{2n}T_n^3 \Gamma_N e^{-2^n/4}/(4M_{N}^3)$ where $\Gamma_N \sim (y^\dagger_D y_D)_{ii} M_{N}/8\pi$ is total decay width of RHN.
Similarly, the thermalization rate can be estimated \cite{Baldes:2021vyz,Huang:2022vkf} as $\Gamma_{\rm th}\approx {\rm ln}(\frac{3 M_{N}^2}{5\pi 2^n\alpha_W T_n^2})\frac{\zeta_3g_{EW}2^n\alpha_{W}^2T_n^3}{4\pi M_{N}^2} $ where $g_{\rm EW}=46, \alpha_W =1/137$. We check that for a typical benchmark point considered in our analysis, $\Gamma_{\rm th} \gg \Gamma_{\rm wash} $ such that all the boosted particles get thermalised quickly via cascade interactions with the SM bath, before inducing any washout effects via inverse decay. Therefore, for our scenario, washout processes are negligible and the baryon asymmetry estimate given by Eq. \eqref{eq:eps} remains valid.

\begin{table}
    \centering
    \begin{tabular}{|c|c|c|c|c|c|c|c|c|c|c|c|c|}
    \hline
       & M & $v_c$  & $T_c$   & $\frac{v_c}{T_c}$ & $\lambda_7 (0)$  & $Y'_{22} (0)$ &  $\lambda_2 (0)$ & $T_n$ & $T_p$ & $\frac{\beta}{{\bf H}_*}$ & $v_J$ &  $\alpha_*$\\
     & (GeV)  & (GeV) & (GeV)  &  & & $\approx Y'_{33} (0)$ & & (GeV) & (GeV) & & & \\
        \hline
       BP1 & $5.54\times10^7$ &  $5.01\times10^7$ & $1.37\times10^7$   & 3.65 & 1.17 & 0.091 & 0.02 & $3.55\times10^6$ & $1.99\times10^6$ & 13.97 & 0.95 & 1.69 \\
       \hline
       BP2 & $7.85\times10^7$ &  $6.97\times10^7$ & $1.93\times10^7$   & 3.61 & 1.18 & 0.090 & 0.02 & $5.00\times10^6$ & $2.76\times10^6$ & 9.71 & 0.95 & 1.76 \\
       \hline
       BP3 & $9.47\times10^7$ &  $8.38\times10^7$ & $2.34\times10^7$   & 3.56 & 1.19 & 0.096 & 0.02 & $5.95\times10^6$ & $3.25\times10^6$ & 3.56 & 0.95 & 1.89 \\
       \hline
       BP4 & $7.95\times10^7$ &  $2.37\times10^7$ & $1.96\times10^7$   & 1.20 & 1.19 & 0.097 & 0.02 & $5.29\times10^6$ & $3.10\times10^6$ & 21.24 & 0.95 & 1.51 \\
       \hline
       BP5 & $2.80\times10^6$ &  $1.54\times10^6$ & $6.57\times10^5$   & 2.36 & 1.06 & 0.086 & 0.02 & $1.70\times10^5$ & $9.78\times10^4$ & 20.81 & 0.95 & 1.75 \\
       \hline
    \end{tabular}
    \caption{Benchmark parameters and other details involved in the GW spectrum calculation of the model.}
    \label{tab1}
\end{table}

\begin{table}[]
    \centering
    \begin{tabular}{|c|c|c|c|c|c|c|c|}
       \hline & $\epsilon_N$  & $T_{\rm RH}$ (GeV) & $T_n$ (GeV)&  $M_{N_{2}} \approx M_{N_3}$ (GeV)& $y_D$ & $\Delta V_{\rm tot} $ (GeV)$^4$  & $M_{\rm DM}$ (MeV)\\
        \hline 
         BP1 & $2.51\times10^{-8}$   & $4.05\times10^6$  & $3.55\times10^6$ &  $3.58\times10^6$ & $4.35\times10^{-5}$ & $1.02\times 10^{28}$  & 12.54\\
        BP2 & $2.60\times 10^{-8}$  & $5.77\times10^6$ & $5.00\times10^6$ & $5.04\times10^6$ & $5.16\times10^{-5}$ & $4.22\times 10^{28}$ & 14.24\\
        BP3 & $2.73\times 10^{-8}$ & $6.98\times10^6$ & $5.95\times10^6$ & $6.43\times10^6$ & $5.83\times10^{-5}$  &$9.05\times 10^{28}$ & 7.86\\
    BP4 &  $2.31\times 10^{-8}$   & $5.87\times10^6$ & $5.29\times10^6$ & $5.50\times10^6$ & $5.39\times 10^{-5}$ & $4.53\times 10^{28}$ & 3.90\\
        BP5 &  $2.56\times 10^{-8}$   & $1.95\times10^5$ & $1.70\times10^5$ & $1.71\times10^5$ & $9.51\times 10^{-6}$ & $5.51\times 10^{22}$ & 4.06\\
        \hline
    \end{tabular}
    \caption{CP asymmetry and other relevant details involved in leptogenesis and dark matter calculation for the model. }
    \label{tab2}
\end{table}

\begin{figure}[h]
    \centering
    \includegraphics[scale=0.5]{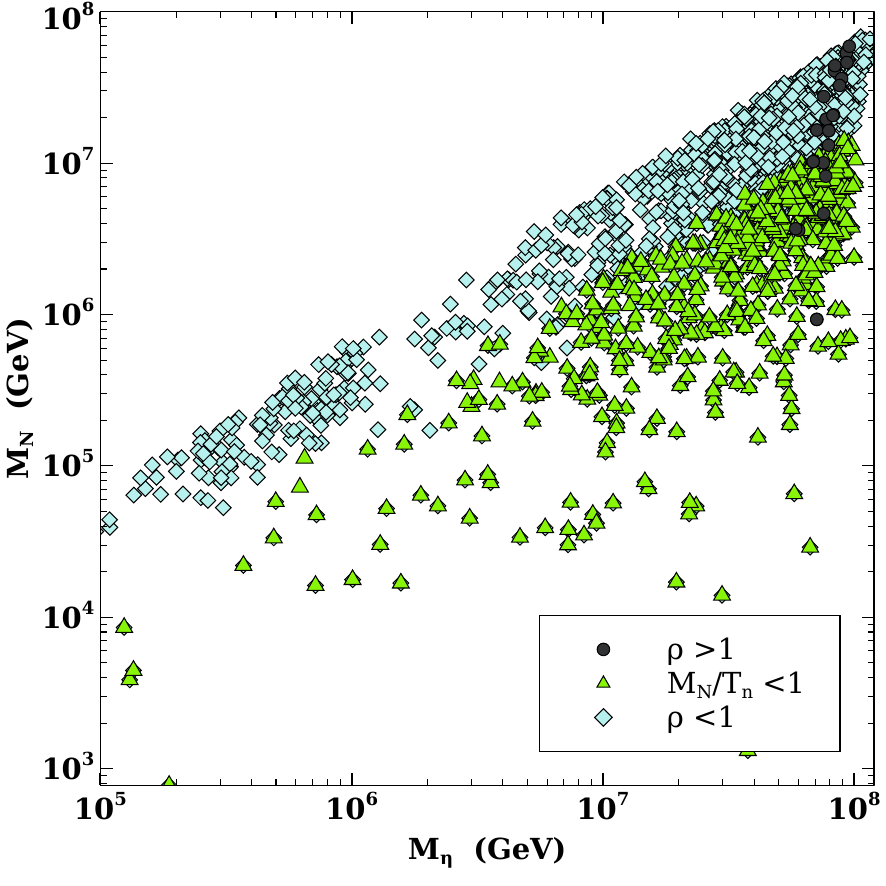}
       \includegraphics[scale=0.5]{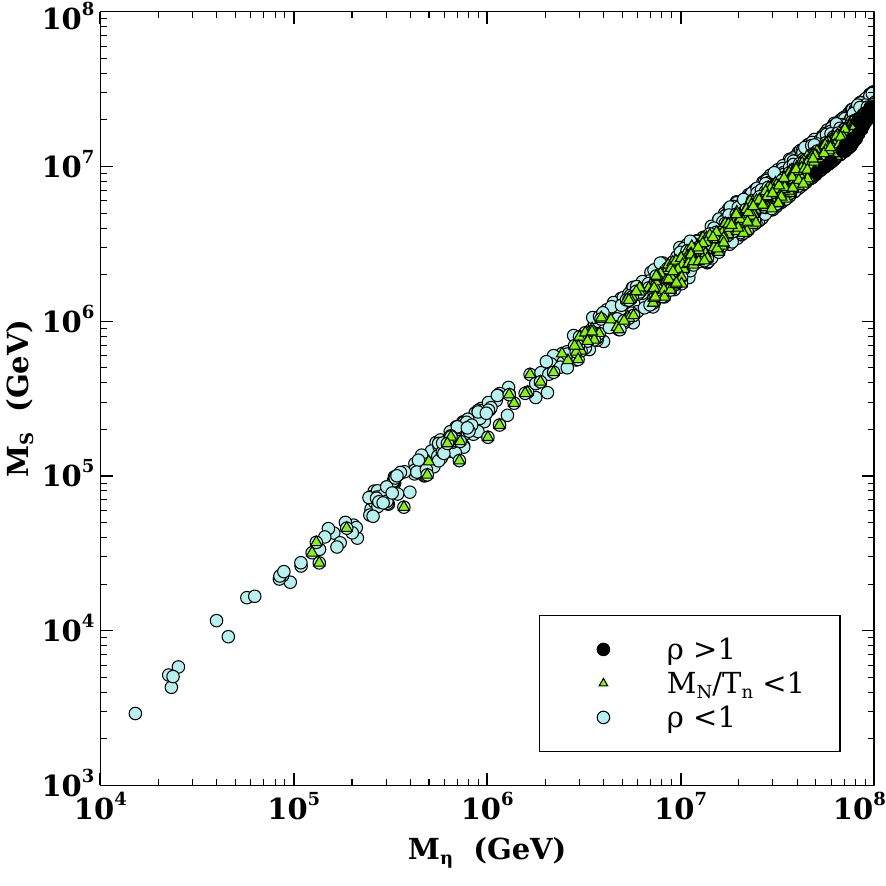}
    \caption{Parameter space in $M_\eta$ versus $M_N$ plane (left panel) and $M_\eta$ versus $M_S$ plane (right panel) consistent with a high scale FOPT showing the SNR with respect to LIGO-VIRGO O3. In this scan, the scale of phase transition (M) is varied from $10^4$ to $10^8$ GeV, $\lambda_7(0)$ is varied from 1 to 3, $Y'_{22}\sim Y'_{33}$ is varied from 0 to 1 and $Y'_{11}$ is varied upto $10^{-2}$.}
    \label{fig:3}
\end{figure}

\begin{figure}[h]
    \centering
    \includegraphics[height=8cm, width=8cm]{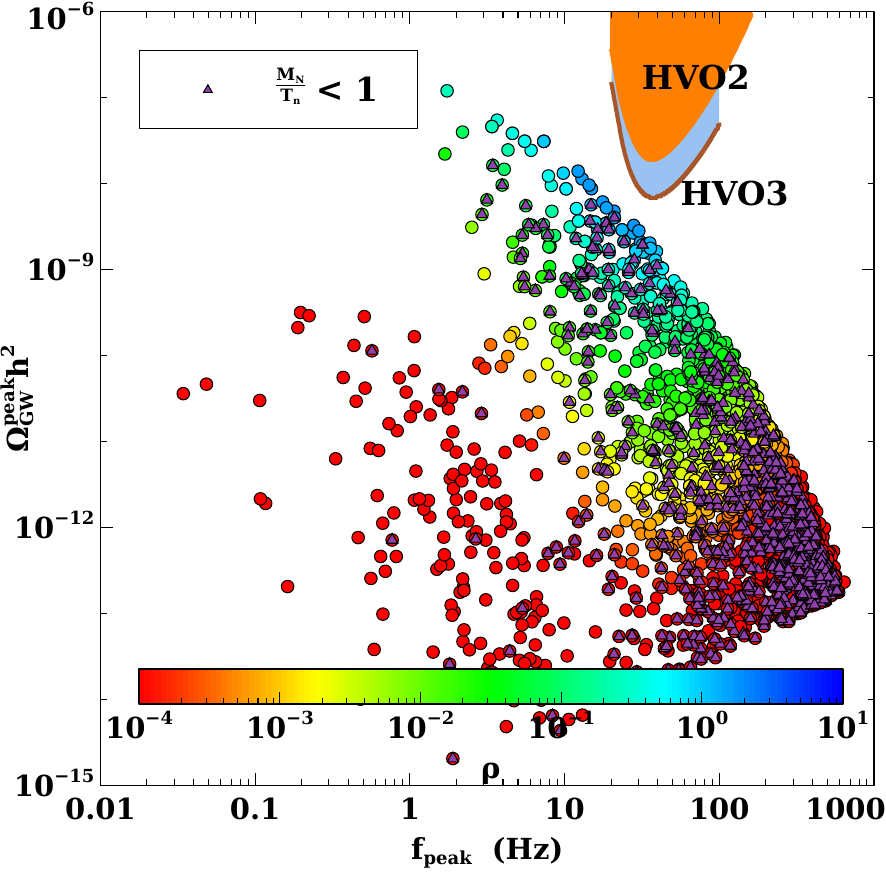}
     \includegraphics[height=8cm, width=8cm]{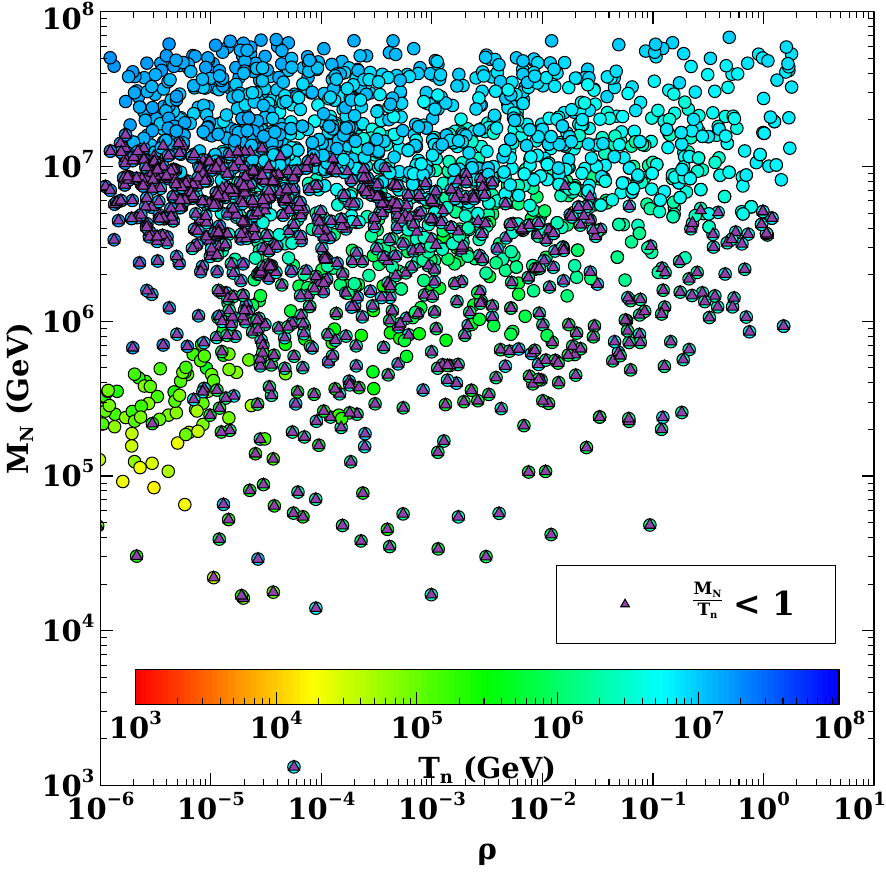}
    \caption{Left panel: GW spectrum peak vs its corresponding Peak frequency with the colour code showing the SNR. Right panel: Heavy RHN mass versus SNR with colour code indicating the corresponding nucleation temperature.}
    \label{fig:4}
\end{figure}

Adopting a conservative approach to apply the GW constraints on the parameter space of our model consistent with a high scale FOPT and leptogenesis, we define the signal-to-noise ratio (SNR) for ongoing LIGO-VIRGO experiment as~\cite{Schmitz:2020syl} 
\begin{equation}
\rho = \sqrt{\tau\,\int_{f_\text{min}}^{f_\text{max}}\,df\,\left[\frac{\Omega_\text{GW}(f)\,h^2}{\Omega_\text{expt}(f)\,h^2}\right]^2}\,, 
\end{equation}
with $\tau$ being the observation time for a particular detector, which we consider to be 1 yr. To register a detection SNR needs to be more than a threshold value $\rho > \rho^{\rm th}$. In this work, we consider $\rho^{\rm th}=1$, such that the parameter space giving rise to SNR $\rho >1$ at LIGO-VIRGO latest run namely O3 is disfavoured due to non-detection of such stochastic GW, as per our conservative selection criteria.

In Fig. \ref{fig:3}, we show the parameter space of our model in terms of physical masses of heavier quasi-degenerate RHNs, singlet and doublet scalar masses in a way similar to the ones in Fig. \ref{fig:1} but without any colour bar. The points corresponding to large SNR namely $\rho >1$ are shown with a different colour. Clearly, a few points in the larger mass range $ \gtrsim 10^7$ GeV with SNR $>1$ get disfavoured. We also indicate the points with $T_n > M_N$ where heavy RHNs can be thermally produced. In such a case, the lepton asymmetry needs to be calculated in a way similar to thermal leptogenesis while solving the relevant Boltzmann equations explicitly. We also calculate the peak amplitudes of GW for the parameter space consistent with FOPT requirements and show them against LIGO-VIRGO sensitivities on the left panel of Fig. \ref{fig:4}. We show the sensitivity curves of run 2 (HVO2) \cite{LIGOScientific:2019vic} as well as run 3 (HVO3) \cite{KAGRA:2021kbb} for comparisons. While the peak frequencies of the points with SNR $>1$ lie outside the sensitivity curves, they can still be detected if we consider the entire GW spectrum. On the right panel of Fig. \ref{fig:4}, we show the heavy quasi-degenerate RHN mass$M_N$ versus SNR with colour code indicating the nucleation temperature. Clearly, some of the points with $T_n \in (10^6-10^7)$ GeV get disfavoured.

Dark matter or the $Z_2$-odd RHN can, in principle, be produced either thermally or non-thermally. For TeV scale FOPT in this model, studied earlier \cite{Borah:2022cdx}, it was found that thermal DM parameter space is almost entirely ruled out due to stringent direct detection bounds. This is precisely due to the constraint on quartic coupling between singlet scalar and the SM Higgs from the requirement of successful electroweak symmetry breaking. This coupling, namely $\lambda_6$ in our notation then decides Higgs portal coupling through which fermion singlet thermal DM typically annihilate. While DM has Yukawa coupling with leptons and $\eta$, Yukawa portal annihilations typically remain sub-dominant especially due to large $\eta$ mass required by FOPT criteria and often require large Yukawa couplings to satisfy relic. Such Yukawa coupling faces tight constraints from neutrino mass as well as charged lepton flavour violation \cite{Toma:2013zsa}. If we go to high scale where some part of the parameter space gets constrained by LIGO-VIRGO data, thermal DM is likely to hit the unitarity limit \cite{Griest:1989wd}. While DM mass need not be same as the scale of FOPT, making it much lighter suppresses the annihilation rates due to heavy mediators in the form of $\eta$ as well as singlet scalar. In addition to mediator suppression, DM coupling with singlet scalar also gets smaller for lighter DM masses. Therefore, thermal DM gets overproduced in such a scenario due to insufficient annihilation rates. While entropy dilution at the end of FOPT could lower this thermal abundance, we find such entropy dilution to be negligible as discussed above, in the context of leptogenesis. Therefore, we consider non-thermal DM which remains feasible for some part of the parameter space under study. 

\begin{figure}[h]
    \centering
    \includegraphics[scale=0.5]{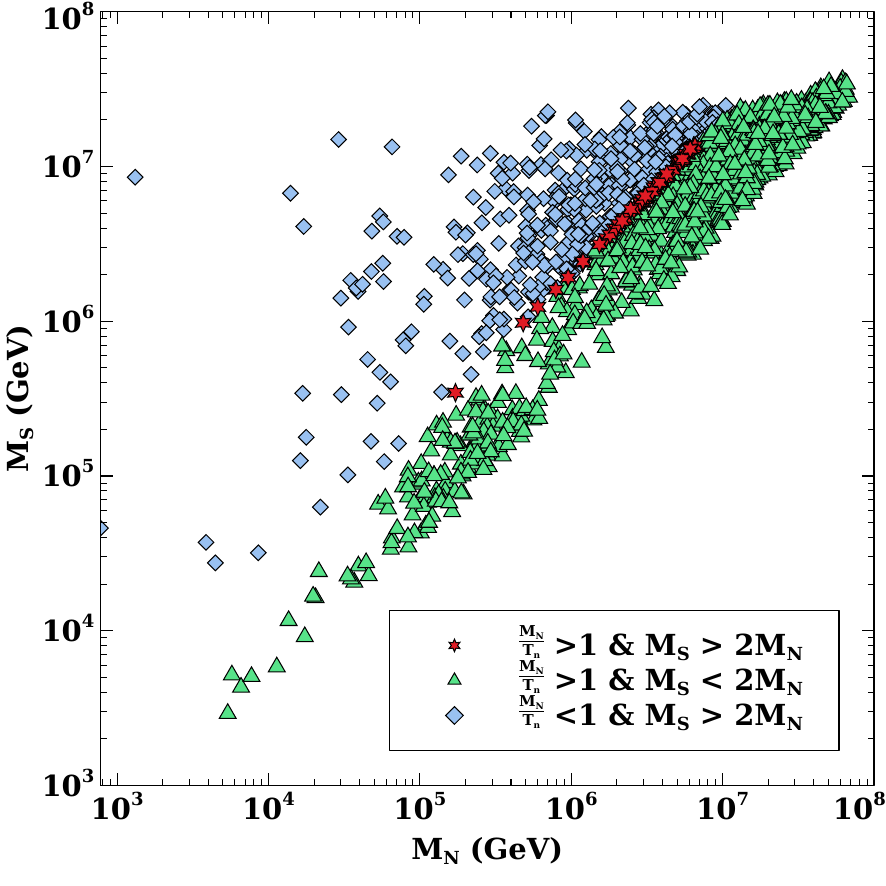}
      \includegraphics[scale=0.5]{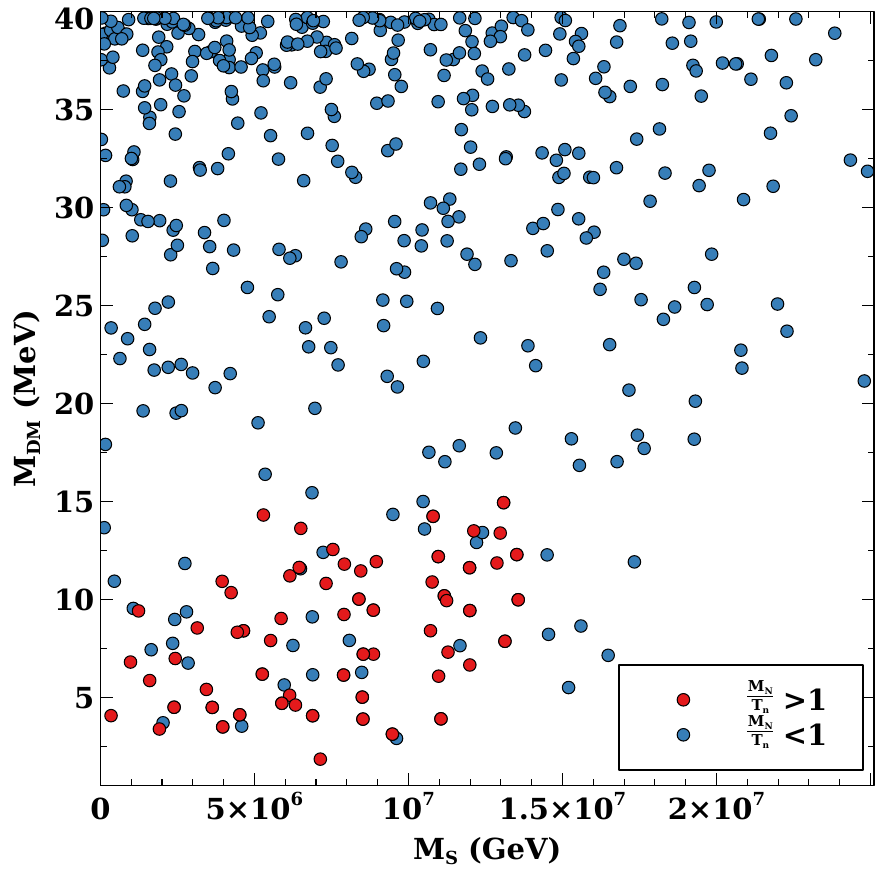}
    \caption{Left panel: The parameter space in $M_N$ vs $M_{S}$ plane where $\diamond$ shaped points are consistent with dark matter relic while $\triangle$ shaped points are not, and $\star$ shaped points satisfy dark matter relic as well as non-thermal leptogenesis criteria. Right panel: The parameter space in $M_S-M_{\rm DM}$ plane which satisfies DM relic criteria. In both the plots, the points labelled as $M_N/T_n >1$ are consistent with non-thermal leptogenesis criteria via mass-gain mechanism.}
    \label{fig:5}
\end{figure}

Such non-thermal or feebly interacting massive particle (FIMP) type DM can be produced dominantly from singlet scalar (S) decay inside the bubble. The corresponding Boltzmann equations for comoving densities of DM and singlet scalar S can be written as
\begin{equation}
    \frac{dY_{\rm DM}}{dz}=\frac{2}{z \bf{H}}  \Gamma_{sDM}  Y_S, \,\,
    \frac{dY_S}{dz}=-\frac{1}{z \bf{H}} ( \Gamma_{sDM} + \Gamma_{sh} + \Gamma_{sN2} +\Gamma_{sN3}) Y_S,
\end{equation}
with $z=M_S/T$ and assuming the relativistic dof to be constant, which is valid at temperatures above electroweak scale. In the second equation, $\Gamma_{sDM}, \Gamma_{sh}, \Gamma_{sN2}, \Gamma_{sN3}$ denote the corresponding decay width of singlet scalar into a pair of DM, SM Higgs, $N_2$, $N_3$ respectively. Similar to $N_{2,3}$, the singlet scalar with mass $M_S > T_n$ is also out-of-equilibrium inside the bubble having a large initial abundance, given by
\begin{align}
    Y_S &= \frac{45}{2\pi^4}\xi(3)\frac{1}{g_*}.
\end{align}
 Since singlet scalar coupling with the SM Higgs is very small, one requires a dominant decay channel of $S$ such that it decays at least before BBN without overproducing DM. Due to $M_S > T_n$ applicable to the entire parameter space, dilution of singlet scalar abundance via annihilation is also suppressed. In the minimal model we are studying, DM overproduction from singlet scalar decay can be avoided only when $M_S > 2 M_{N_{2,3}} \sim 2 M_N$ keeping the branching ratio BR$(S \rightarrow {\rm DM \, DM})$ small. We have chosen our benchmark points in table \ref{tab1}, \ref{tab2} such that this condition is satisfied. The corresponding DM masses, consistent with correct relic abundance, are shown in the last column of table \ref{tab2}. On the left panel of Fig. \ref{fig:5}, we show the parameter space in $M_S-M_N$ plane while indicating the region consistent with DM relic criteria ($\diamond$ shaped points) and both DM and non-thermal leptogenesis criteria ($\star$ shaped points). On the right panel of the same figure, we show the mass of DM as a function of singlet scalar mass, which satisfy correct DM relic criteria (blue coloured points) and both DM and non-thermal leptogenesis criteria (red coloured points). We also check the contribution of SM Higgs decay to FIMP DM relic and find it to be negligible, due to the smallness of singlet-SM Higgs mixing. Thus, incorporating DM relic criteria puts tight constraints on the model parameter space.

\section{Conclusion}
\label{sec:conclude}

We have studied the possibility of a high scale first order phase transition such that the resulting gravitational wave spectrum can be constrained by ongoing GW experiments like LIGO-VIRGO. The FOPT not only leads to such stochastic GW spectrum but also triggers the production of baryon asymmetry via leptogenesis as well as dark matter via the mass-gain mechanism. Due to the high scale nature of the FOPT, the DM is favourably in the non-thermal or FIMP ballpark. The combined criteria of successful leptogenesis and DM relic constrain the model parameter space as well as the mass spectrum of BSM particles. Some part of the parameter space also lead to a large signal to noise ratio at LVK experiments and hence can be disfavoured in a conservative manner, owing to non-observations of such stochastic spectrum. While it is interesting to constrain a realistic particle physics setup explaining the origin of baryon asymmetry and dark matter from ongoing GW experiments, more data from ongoing as well as future experiments will shed more light into the remaining parameter space. Since a high scale FOPT setup typically remains out of reach of direct experimental search, such indirect constraints and complementary signatures remain worth exploring.

\section*{Acknowledgements}
The work of DB is supported by SERB, Government of India grant MTR/2022/000575. 
\appendix

\section{Finite temperature correction to potential}
\label{appen1}
The finite-temperature one-loop potential is given by
\begin{align}
V_{\rm tot} = V_{\rm tree} + V_{\rm CW} +V_{\rm th}.
\end{align}
While the tree level potential is given by Eq. \eqref{eq:tree potential}, the Coleman-Weinberg potential~\cite{Coleman:1973jx} with $\overline{\rm DR}$ regularisation is given by
\begin{align}
V_{\rm CW} = \sum_i (-)^{n_{f}} \frac{n_i}{64\pi^2} m_i^4 (\phi) \left(\log\left(\frac{m_i^2 (\phi)}{\mu^2} \right)-\frac{3}{2} \right),
\end{align}
where suffix $i$ represents particle species, and $n_i,~m_i (\phi)$ are the degrees of freedom (dof) and field dependent masses of $i$'th particle. In addition, $\mu$ is the renormalisation scale, and $(-)^{n_f}$ is $+1$ for bosons and $-1$ for fermions, respectively. Since we are tracking the singlet scalar field for FOPT, we consider its vacuum expectation value (VEV), denoted by $M$ as the renormalisation scale as $\mu = M = \langle S \rangle$. We denote the singlet scalar as $S=(\phi+M)/\sqrt{2}$.
The field dependent masses associated with phase transition with dof are

\begin{equation}
    m_\eta^2=\lambda_7 \phi^2/2 \; (n_\eta = 4), \quad m_s^2=3\lambda_s \phi^2 \; (n_s =1), \quad m_{y_i}^2=2 y_i^2 \phi^2 \; (n_y=2).
\end{equation}

The thermal contributions to the effective potential are given by

\begin{align}
V_{\rm th} = \sum_i \left(\frac{n_{\rm B_i}}{2\pi^2}T^4 J_B \left[\frac{m_{\rm B_i}}{T}\right] - \frac{n_{\rm F_{i}}}{2\pi^2}T^4 J_F \left[\frac{m_{\rm F_{i}}}{T}\right]\right),
\end{align}

where $n_{B_i}$ and $n_{F_i}$ denote the dof of the bosonic and fermionic particles, respectively. The $J_B$ and $J_F$ functions are defined by following functions:

\begin{align}
&J_B(x) =\int^\infty_0 dz \, z^2 \log\left[1-e^{-\sqrt{z^2+x^2}}\right] \label{eq:J_B},\\
&J_F(x) =   \int^\infty_0 dz \, z^2 \log\left[1+e^{-\sqrt{z^2+x^2}}\right].
\end{align}

In thermal potential, we also consider the contribution from daisy diagrams \cite{Fendley:1987ef,Parwani:1991gq,Arnold:1992rz}. Considering Arnold-Espinosa method \cite{Arnold:1992rz}, the thermal potential with the daisy correction can be written as

\begin{align}
    V_T(\phi,T) &= V_{\rm th} + V_{\rm daisy}(\phi,T), \\
    V_{\rm daisy}(\phi,T) &= -\sum_i \frac{g_i T}{12\pi}\left[ m^3_i(\phi,T) - m^3_i(\phi) \right]. \nonumber
\end{align}

Denoting $m^2_i(\phi,T)=m^2_i(\phi) + \Pi_i(T)$, the relevant thermal masses can be written as \cite{Cline:2008hr}

\begin{align*}
\Pi_\eta(T)=(\frac{g_2^2}{8}+\frac{g_1^2+g_2^2}{16}+\frac{\lambda_2}{2}+\frac{\lambda_3+\lambda_4}{12})T^2, \\
\Pi_s(T)=(\frac{\lambda_s}{4}+\frac{\lambda_6}{3}+\frac{\lambda_7}{3}+\frac{y_1^2}{8}+\frac{y_2^2}{8}+\frac{y_2^2}{8})T^2.
\end{align*}

\section{Action calculation through fitting of the finite temperature potential}
\label{appen3}

In general, the tunneling rate per unit time per unit volume between two minima separated by a barrier under an arbitrary potential can be written as

\begin{equation}
    \Gamma_V = \mathcal{A} e^{-B}
\end{equation}

where, the coefficients $\mathcal{A}$ and B depend on the potential. In Euclidean space or imaginary time formalism, the coefficient B is $S_3/T$ with $S_3$ being the Euclidean action at finite temperature in three dimensions. As described in \cite{Linde:1980tt}, the dimensional estimation of pre-exponential factor $\mathcal{A}$ is $T^4\left( \frac{S_3(T)}{2\pi T}\right)^{3/2}$ considering three zero modes of solution for thermal tunneling. So, the probability of thermal tunneling per unit time per unit volume can be written as

\begin{align}
\Gamma (T) = \mathcal{A}(T) e^{-S_3(T)/T},
\end{align}

where $\mathcal{A}(T)\sim T^4 \left( \frac{S_3(T)}{2\pi T}\right)^{3/2}$.  In order to calculate the action $S_3$, we need to solve the O(3) symmetric equation of motion i.e. bounce equation

\begin{equation}
    \frac{d^2 \phi}{d r^2}+ \frac{2}{r}\frac{d \phi}{d r}=\frac{\partial V(\phi,T)}{\partial \phi}
\end{equation}

satisfying the boundary conditions

\begin{align}
    \phi(r\rightarrow \infty)=\phi_{\rm false} ~~~ {\rm and} ~~~ \left. \frac{d \phi}{d r}\right|_{r=0}=0.
\end{align}

The action corresponding to the solution of the above equation is
\begin{equation}
    S_3(T)= 4\pi \int_0^\infty r^2 dr \left [ \frac{1}{2} \left(\frac{d \phi}{d r}\right)^2 + V(\phi(r),T) \right ].
\end{equation}
Here, we used a fitting method to calculate the action. We fitted the effective potential to a generic potential for which the action calculations are done in semi-analytical way \cite{Adams:1993zs}. The generic quartic and logarithmic potential is 
\begin{equation}
    V(\phi)= (2A-B) \mu^2 \phi^2 -A\phi^4 + B\phi^4 {\rm ln} \left(\frac{\phi^2}{\mu^2} \right).
\end{equation}

\begin{figure}[h]
    \centering
    \includegraphics[scale=0.5]{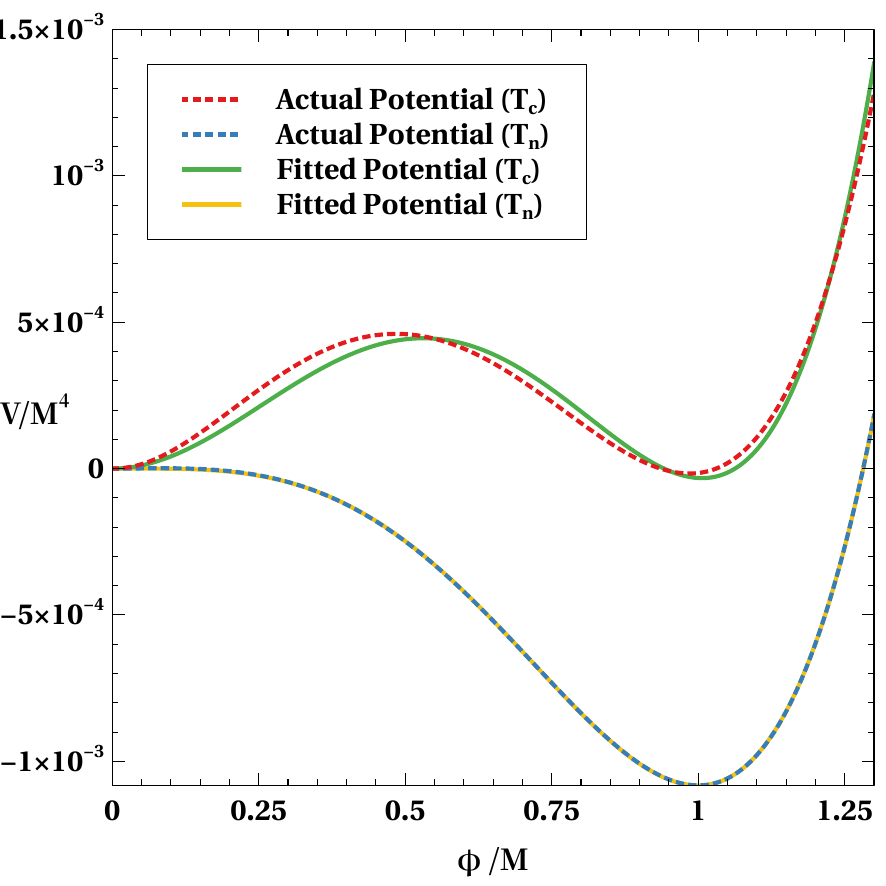}
    \includegraphics[scale=0.5]{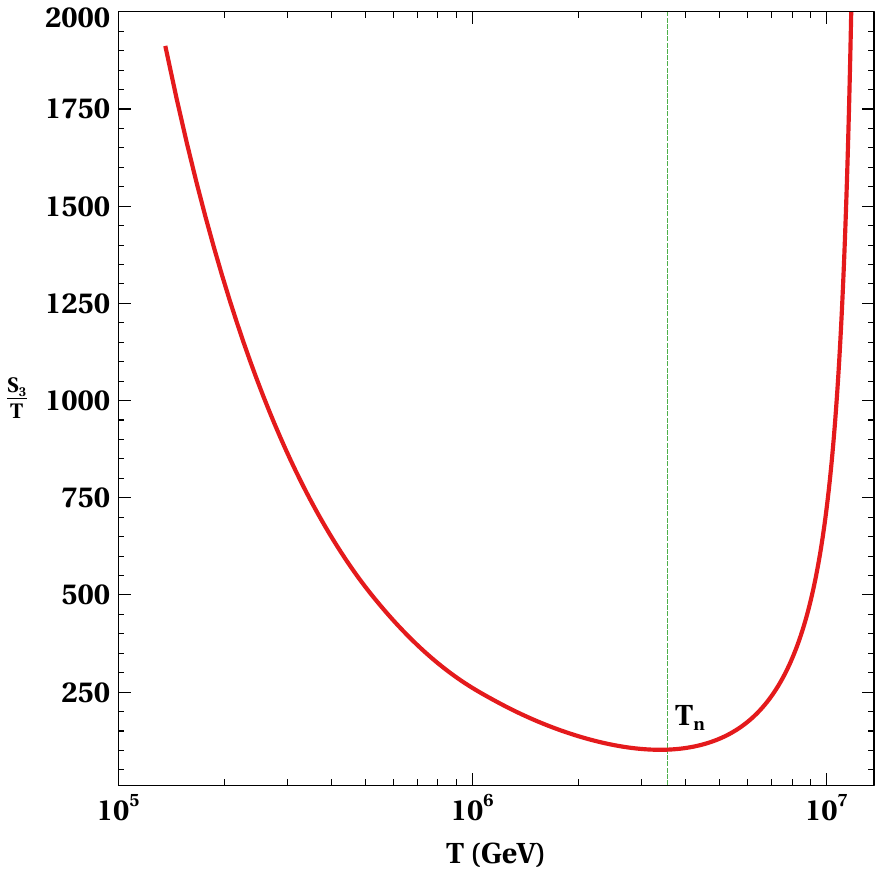}
    \caption{Left Panel: Comparison between actual potential and fitted generic potential at critical temperature $T_c$ and nucleation temperature $T_n$ for BP1. Right Panel: $S_3/T$ as a function of temperature for BP1.}
    \label{fitting}
\end{figure}

We show that the actual potential and the fitted generic potential overlap quite well, depicted in left panel of Fig. \ref{fitting} for BP1. The bounce equation can be simplified with dependence on just one parameter using scaling transformation ( $\Phi=\phi/\mu$ and $a=2 \mu \sqrt{B}r$) as
\begin{equation}
    \frac{d^2 \Phi}{d a^2}+ \frac{2}{a}\frac{d \Phi}{d a}=\delta(\Phi-\Phi^3)+\Phi^3 {\rm ln} \Phi^2
\end{equation}
where $\delta \equiv \frac{2A-B}{2B}$. As discussed in \cite{Adams:1993zs}, the desired Euclidean action is calculated in semi-analytical fashion and can be written as 
\begin{equation}
    S_3=\frac{16 \pi \sigma I^3}{3(1-2\delta)^2}\Bigl(\frac{2}{B}\Bigl)^{1/2}(2\delta)^{n_\mu}\Bigl\{1+\mu_1\delta+\mu_2\delta^2+\mu_3\delta^3\Bigl\}
\end{equation}
where, I=0.4199, $n_\mu=0.557$, $\mu_1=4.2719$, $\mu_2=-14.5908$ and $\mu_3=12.0940$. In our analysis, we fitted the effective potential using the above generic expression of potential for a wide range of temperatures such that we have a profile of action over temperature shown in right panel of Fig. \ref{fitting} for BP1. That plays an important in tunneling rate and $\beta/\mathcal{H}$ calculations.

\section{Renormalisation Group Evolution Equations}
\label{appen2}

The relevant RGE equations for the model parameters are \cite{Bhattacharya:2019tqq}

    \begin{equation*}
        \frac{d \lambda_s}{dt}=\frac{1}{16\pi^2}(20\lambda_s^2+ 2\lambda_6^2+2\lambda_7^2+8\lambda_s {\rm Tr}[Y'^{\dagger} Y']-{\rm Tr}[Y'^{\dagger} Y'Y'^{\dagger} Y']),
    \end{equation*}
    \begin{equation*}
    \frac{d \lambda_2}{dt}=\frac{1}{16\pi^2}(12\lambda_2^2+2\lambda_7^2+3g_1^2/4 +9g_2^2/4 +3g_1^2g_2^2/2),
    \end{equation*}
   \begin{equation*}
   \frac{d \lambda_7}{dt}=\frac{1}{16\pi^2}(4\lambda_7^2+6\lambda_2 \lambda_7+8\lambda_s\lambda_7 +4\lambda_7 {\rm Tr}[Y'^{\dagger} Y']),
   \end{equation*}
   \begin{equation*}
   \frac{d \lambda_6}{dt}=\frac{1}{16\pi^2}(4\lambda_6^2+6\lambda_6 y_t^2+8\lambda_s\lambda_6 +4\lambda_6 {\rm Tr}[Y'^{\dagger} Y']),
   \end{equation*}
   \begin{equation*}
   \frac{d Y'}{dt}=\frac{1}{16\pi^2}(4Y'^3+2Y' {\rm Tr}[Y'^{\dagger} Y']),
   \end{equation*}
   \begin{equation*}
   \frac{d g_1}{dt}=\frac{1}{16\pi^2}(7g_1^3),
   \end{equation*}
   \begin{equation*}
   \frac{d g_2}{dt}=\frac{1}{16\pi^2}(-3g_2^3),
   \end{equation*}
   \begin{equation*}
   \frac{d y_t}{dt}=\frac{1}{16\pi^2}(9y_t^3/2-y_t(17g_1^2/12+9g_2^2/4)).
   \end{equation*}


\end{document}